\documentclass[aps,prl,twocolumn,groupedaddress]{revtex4}

\usepackage{amsmath,amssymb,color}
\usepackage[usenames,dvipsnames]{xcolor}
\usepackage{multirow} 
\usepackage{hyperref}

\newcommand{\noperp}{{\phantom \perp\!}}

\begin{document}


\title{Nonlinear Partially Massless from Massive Gravity?}


\author{S.~Deser}
\email[]{deser@brandeis.edu}
\affiliation{Lauritsen Lab, Caltech, Pasadena CA 91125 and Physics Department, Brandeis University, Waltham, MA 02454, USA}

\author{M.~Sandora}
\email[]{sandora@ms.physics.ucdavis.edu}
\affiliation{Department of Physics, University of California, Davis, CA 95616, USA}

\author{A.~Waldron}
\email[]{wally@math.ucdavis.edu}
\affiliation{Department of Mathematics, University of California, Davis, CA 95616, USA}


\date{\today}

\begin{abstract}
We show that  consistent nonlinear Partially Massless models cannot be obtained 
starting from ``$f$-$g$'' massive gravity, with ``$f$'' the embedding de Sitter space. The 
obstruction, which is  also the  source of $f$-$g$ acausality, is the very same
 fifth constraint that removes the notorious sixth ghost excitation. Here, however, it blocks
extension of the mass to cosmological background  tuned gauge invariance that took
away the helicity zero mode at linear level. Separately, our methods allow us to almost complete  the class of acausal $f$-$g$ models.  

\end{abstract}


\maketitle

\section{Introduction}

The, by now well-appreciated, fact~\cite{PM} that de Sitter~(dS) space representations allow for novel gauge invariances of otherwise massive free flat space higher ($s\geq2$) spins has led to hopes for extensions of these partially massless (PM) models into the nonlinear realm. The lowest spin, and most interesting, extension is that of spin 2 PM to ``PM gravity'' (PMG). Unfortunately, that hope has already been excluded in several  contexts. Firstly, a comprehensive perturbative study of higher spin extensions~\cite{Zinoviev} has noted (without giving details), that an obstruction indeed arises at quartic order (cubic extensions, being simply Noether current couplings, are always trivially allowed). A different approach, based on the observation~\cite{Maldacena} that conformal, Weyl, gravity kinematically describes both $s=2$ PM and Einstein graviton modes about dS vacuum, led to a recent search based on suitably truncating the Weyl model~\cite{DJ}. Here too, an obstruction was encountered beyond cubic order. Separately, a different tack has been taken by two groups~\cite{PMmg,PMmg1}, based on their currently popular massive gravity models (for a review, see~\cite{HB}). These are ({\it ab initio} nonlinear) Einstein gravities, but with very special mass terms involving a preferred background, ``$f$'', metric,  that preserve the five degree of freedom (DoF) content of linear Fierz-Pauli (FP) massive $s=2$. Taking this background to be a suitably ``tuned'' dS, they hope to define a PMG~\footnote{While it is not clear to us if the above two models coincide, this is irrelevant: we will show that there are is no consistent PMG. Of course if their purportedly unique PMG candidate theories are different, then absence of PMG is already proven by contradiction!}. 
Our purpose here is to show that this avenue is unfortunately also blocked. We will find that the very dS gauge invariance required to eliminate the massive model's helicity-0 mode cannot be implemented at nonlinear level: it would have to turn that fifth constraint into a Bianchi identity, thereby removing helicity-0 at the tuned point. But this is obstructed precisely due to the same set of its terms that lead to the massive model's becoming acausal~\cite{DWa}. The irony is again that the very special set of mass terms that are the solution to avoiding the ancient Boulware-Deser~\cite{BD} sixth DoF ghost catastrophe,
now become part of the problem. Indeed, a byproduct of the present work will be to extend the set of acausal mass terms in the massive theory, leaving only  one (unlikely) window there--and none for PMG.

\section{The Model}

We begin with the, most general, five-parameter, family of 
$f$-$g$ massive GR actions known to have five (rather than six) DoF~\cite{HR}; their field equations are:
\begin{equation}\label{eom}
{\cal G}_{\mu\nu}:=G_{\mu\nu}-\Lambda g_{\mu\nu} -\sum_{i=1}^3 \mu_i \tau_{\mu\nu}^{(i)}=0\, ,
\end{equation}
where~\footnote{The mass terms $\tau_{\mu\nu}^{(i)}$ are generated by the 
compact expression~\cite{Deffayet}$$\ \ \  \: \: \varepsilon_\mu{}^{\mu_1\mu_2\mu_{3}}\varepsilon_{\nu m_1m_2 m_{3}} f_{\mu_1}{}^{m_1}\cdots f_{\mu_i}{}^{m_i}e_{\mu_{i+1}}{}^{m_{i+1}}\cdots e_{\mu_{3}}{}^{m_{3}}\, .$$}
\begin{eqnarray}
\tau^{(1)}_{\mu\nu}&:=&f_{\mu\nu} - g_{\mu\nu} f\, ,\nonumber\\[2mm]
\tau^{(2)}_{\mu\nu}&:=&2 \big(f_{\mu\rho}-g_{\mu\rho} f\big) f^\rho_\nu + g_{\mu\nu}\big(f^2 - f_{\rho}^{\sigma} f^{\rho}_{\sigma} \big)\, ,\nonumber\\[2mm]
\tau^{(3)}_{\mu\nu}&:=&6 \big(f_{\mu\rho}-g_{\mu\rho} f\big) f^\rho_\sigma f^\sigma_\nu  
+ 3f_{\mu\nu}\big(f^2 - f_{\rho}^{\sigma} f^{\rho}_{\sigma}  \big)\nonumber\\[1mm]
&-&g_{\mu\nu}\big(f^3-3 f f_{\rho}^{\sigma} f^{\rho}_{\sigma} + 2 f_{\rho}^{\sigma} f^{\eta}_{\sigma} f_\eta^\rho \big)\, .\nonumber
\end{eqnarray}
The metric $g_{\mu\nu}$ is the only  dynamical field and $G_{\mu\nu}$ is its Einstein tensor. The last of the five parameters
$(\Lambda,\mu_1,\mu_2,\mu_3,\bar\Lambda)$ is encoded in the curvature of the non-dynamical vierbein $f_\mu{}^m$:
\begin{equation}\label{backR}
\bar R_{\mu\nu}{}^{mn}:= \bar W_{\mu\nu}{}^{mn} + \frac{2\bar\Lambda}3\,  f_{[\mu}{}^m f_{\nu]}{}^n\, .
\end{equation}
We are primarily interested in the case where the background metric $\bar g_{\mu\nu} :=f_\mu{}^m f_{\nu m}$ is constant curvature (Eq.~(\ref{backR}) with vanishing Weyl tensor $\bar W_{\mu\nu}{}^{mn}$)
but  our results also apply to the more general case of Einstein backgrounds~\footnote{Linear PM  can indeed propagate in Einstein backgrounds~\cite{DNW}; geometrically this may be viewed as the obstruction to a conformally Bach-flat (conformal gravity) metric being conformally Einstein~\cite{DJ}.}.
 All indices are raised and lowered with the dynamical metric
and its vierbein $e_\mu{}^m$ so that (perhaps somewhat confusingly for bimetric theorists)
\begin{equation}\label{f}
f_{\mu\nu}:=f_\mu{}^m e_{\nu m}\, .
\end{equation}
Moreover we require~\footnote{Here, we take Eq.~(\ref{f}) as part of the definition
of the theory, although, generically it can be derived from the equations of motion and in turn an underlying action principle~\cite{Deffayet}. Ruling out non-generic  vierbeine solutions not subject to Eq.(\ref{fsymm})
is akin to the invertibility requirement placed on them and the metric.}
\begin{equation}\label{fsymm}
f_{\mu\nu}=f_{\nu\mu}\, ,
\end{equation}
which gives six independent relations that, along with $g_{\mu\nu}=e_{\mu}{}^m e_{\nu m}$, 
determine the sixteen components of the vierbein $e_\mu{}^m$ in terms of the  ten dynamical metric components. 
The equations of motion have been proven to propagate five DoF for generic parameter values in~\cite{HR}. A simple covariant proof for the special case $\mu_3=0$
has been given in~\cite{Deffayet} (see also~\cite{DWa}). Before proceeding to a  covariant constraint analysis, let us  review the appearance of the PM model in the
linearized theory. 

\section{Linear PM}
\label{lPM}

To linearize the equation of motion~(\ref{eom}) about a background Einstein metric~$\bar g_{\mu\nu}$ we call
$$
h_{\mu\nu}:=g_{\mu\nu}-\bar g_{\mu\nu}\ \Longrightarrow\  f_{\mu\nu}\approx \ \bar g_{\mu\nu}+\frac 12 \, h_{\mu\nu}\, .
$$
Noting that~\footnote{For the remainder of this Section (only) we raise and lower indices with the background metric $\bar g_{\mu\nu}$.}
\begin{eqnarray*}
G_{\mu\nu}\;\;\;&\approx& \:\  \bar \Lambda\,  \bar g_{\mu\nu} + G^L_{\mu\nu}\, ,\nonumber\\[2mm]
(\delta_\mu^\rho - h^\rho_\mu) \sum_{i=1}^3 \mu_i \tau_{\rho\nu}^{(i)}\!&\approx&\!\!
 -\ \! 3 \, (\mu_1-2\mu_2+2\mu_3) \ \bar g_{\mu\nu}\\&-&\  \frac12\, \big(\mu_1 -4\mu_2 +6\mu_3) \, \big[h_{\mu\nu}-\bar g_{\mu\nu} h\big]\, ,
\end{eqnarray*}
we obtain the linearized equation of motion 
\begin{eqnarray}\label{leom}
G_{\mu\nu}^L -\bar\Lambda h_{\mu\nu}&\approx&
\big(\Lambda-\bar\Lambda -\ 3 \mu_1+6\mu_2-6\mu_3) \, \bar g_{\mu\nu}\label{leom}\\&-&\frac12\, \big(\mu_1 -4\mu_2 +6\mu_3) \, \big[h_{\mu\nu}-\bar g_{\mu\nu} h\big]\, .\nonumber
\\ \nonumber
\end{eqnarray}
For models obeying 
$
\Lambda-\bar\Lambda -\ 3 \mu_1+6\mu_2-6\mu_3 = 0\, ,
$
the constant term vanishes and $g_{\mu\nu}=\bar g_{\mu\nu}$ is a solution. 
We thus identify the FP mass
$$
m^2=-\mu_1 +4\mu_2-6\mu_3 \, .
$$
The PM tuning is $m^2=\frac{2\bar \Lambda}3$ at which value the linearized model enjoys the gauge invariance
$$
\delta h_{\mu\nu}=\big(\bar \nabla_\mu\partial_\nu +\frac {\bar \Lambda}3\,  \bar g_{\mu\nu}\big) \alpha\, .
$$
This, along with the vector constraint $\bar\nabla.h_\nu-\bar\nabla_\mu h=0$ following from the divergence  of the linearized equation of motion ${\cal G}^L_{\mu\nu}=0$ determined by~(\ref{leom}),
reduces the ten components of the dynamical field $h_{\mu\nu}$ to four propagating ones. 
Gauge invariances are associated with Bianchi identities; in our case, with
$$
\bar \nabla^\mu \bar \nabla^\nu {\cal G}^L_{\mu\nu} + \frac{\bar \Lambda}3 \,\bar g^{\mu\nu}  {\cal G}^L_{\mu\nu}\equiv0\, .
$$
Our main goal is to search for a non-linear version of this Bianchi identity.

\section{The Fifth Constraint and Putative PM Model}

Returning to the non-linear equation of motion~(\ref{eom}) and taking its divergence, we immediately uncover a vector constraint
\begin{equation}\label{vector}
0={\cal C}_\nu:=\nabla^\mu{\cal G}_{\mu\nu}=-\sum_{i=1}^3 \mu_i \nabla^\mu\tau_{\mu\nu}^{(i)}\, .
\end{equation}
The right hand side was obtained using the Bianchi identity for the Einstein tensor $G_{\mu\nu}$ and contains at most one derivative on the 
dynamical metric. Presently, we will need explicit expressions for the right hand side of~(\ref{vector}) but first present an ``index-free'' sketch
of how a fifth, scalar constraint arises. In particular, we focus on whether this constraint can morph into a Bianchi identity. Our scheme is to organize the scalar constraint 
in powers of the background vierbein~$f$ and derivatives of the dynamical metric.

Since the non-linear mass terms $\tau^{(i)}$ depend algebraically on $f$ and $g$, their covariant derivatives appearing in the vector constraint~(\ref{vector}),
take the form $f^{i-1} \nabla f$. Of course $\bar \nabla f \equiv 0$, so $\nabla f$ measures the difference between the Levi-Civita connections of $e$ and $f$, or in other
words the contorsion $K$ (see Eq.~(\ref{contorsion}) below) which counts as one metric derivative. Hence the vector constraint takes the form 
$$
0=\mu_1 K f + \mu_2 f K f + \mu_3 f^2 K f\, .
$$ 
Multiplying this expression by $f^{-1}$ and taking a further divergence yields
\begin{equation}\label{divC}
0=\mu_1 \nabla K + \mu_2 \nabla (f K) + \mu_3 \nabla(f^2 K)\, .
\end{equation}
This scalar relation involves two derivatives on the dynamical metric so is not a constraint.
However, contracting the field equation ${\cal G}_{\mu\nu}$ on either the metric or $f_{\mu\nu}$ (and powers thereof) also produces a
scalar depending on two metric derivatives. In particular, the Riemann tensor $R(g)$ of the metric $g$ can be expressed 
in terms of its $\bar g$ counterpart and contorsions. Thus, using Eq.~(\ref{backR}), the Einstein tensor can be expanded as
$$
G(g)=\bar \Lambda f^2 + \bar W+ \nabla K + K^2\, . 
$$
Thus the contracted field equation yields
\begin{widetext}
\begin{equation}\label{snooker}
\begin{array}{ccccccccccccccccc}
\mu_1 {\cal G}  +   \mu_2 f {\cal G}  +  \mu_3 f^2 {\cal G} 
&=& {\color{green}\mu_1 \bar\Lambda f^2} &+& {\color{JungleGreen}\mu_1  \bar W} &+& \mu_1 \nabla K &+& \mu_1 K^2 &+&{\color{red} \mu_1 \Lambda}&+& {\color{blue}\mu_1^2 f} &+& {\color{green}\mu_1 \mu_2 f^2} &+& {\color{pink} \mu_1\mu_3 f^3}\\[2mm]
&+& {\color{pink}\mu_2 \bar\Lambda f^3} &+& {\color{JungleGreen}\mu_2 f\bar W} &+& \mu_2 f\nabla K &+& \mu_2 f K^2 &+&{\color{blue}\mu_2  \Lambda f}&+& {\color{green}\mu_2\mu_1 f^2} &+&{\color{pink}\mu_2^2f^3} &+& {\color{brown}\mu_2\mu_3 f^4}\\[2mm]
&+& {\color{brown}\mu_3 \bar\Lambda f^4} &+& {\color{JungleGreen}\mu_3 f^2\bar W} &+& \mu_3 f^2\nabla K &+& \mu_3 f^2 K^2 &+&{\color{green}\mu_3 \Lambda f^2}&+& {\color{pink}\mu_3\mu_1 f^3} &+&{\color{brown}\mu_3 \mu_2 f^4} &+& {\color{Magenta}\mu_3^2 f^5}\, .\\
\end{array}
\end{equation}
\vspace{2mm}
\end{widetext}
There are two criteria we can place on this relation: (i)~For a fifth covariant constraint to exist, the double derivative metric terms in the third column on the right hand side must cancel once one employs  the double divergence of the field equation given in Eq.~(\ref{divC}). (ii) For a Bianchi identity signaling PM, all remaining terms must cancel (these are color-coded for easy reading).
For models with non-vanishing $(\mu_1,\mu_2)$ and $\mu_3=0$, criterion~(i) has been proven to hold~\cite{Deffayet}. The case $\mu_3\neq 0$ is still an open question, but will soon turn out to be irrelevant
for our PM considerations. We thus turn to the second, PM, criterion.

To study criterion (ii), we first examine terms algebraic in~$f$ order by order. At order zero (red), there is only a single term forcing the parameter-constraint
$$
\mu_1 \Lambda = 0\, ,
$$
while at order one (blue) there are two terms: $\mu_1^2 f + \mu_2 \Lambda f$. In the case $\Lambda\neq 0$ we are forced to set $\mu_1=0$ and in turn $\mu_2=0$.
Coupled with the fact that there is only a single term $\mu_3^2 f^5$ at order five (magenta), which imposes $$\mu_3=0$$ (the tensor structure $f^5$ is generically non-vanishing~\footnote{This point deserves some clarification: There is in fact a single choice of contraction of two $f$'s on the mass term~$\tau_{\mu\nu}^{(3)}$ that vanishes, namely $[f^\mu_\rho f^{\rho\nu}-\frac14\, f f^{\mu\nu}] \tau_{\mu\nu}^{(3)}\equiv 0$ (this is easily verified by diagonalizing the symmetric matrix $f_{\mu\nu}$). In that case one must consider the terms in Eq.~(\ref{snooker}) quartic in $f$, namely $\mu_2\mu_3 f^{\mu\nu}\tau^{(3)}_{\mu\nu}$, $\mu_2\mu_3 [f^\mu_\rho f^{\rho\nu}-\frac14\, f f^{\mu\nu}] \tau_{\mu\nu}^{(2)}$ and $\mu_3\bar\Lambda 
[f^\mu_\rho f^{\rho\nu}- \frac14\,f f^{\mu\nu}] \bar G_{\mu\nu}$ where $\bar G_{\mu\nu}:=f_\mu^\rho f_{\rho\nu}-ff_{\mu\nu}-\frac12  g_{\mu\nu}(f^\rho_\sigma f_\rho^\sigma-f^2)$. The only combination of these three tensor structures that vanishes is $[f^\mu_\rho f^{\rho\nu}-\frac14\, f f^{\mu\nu}] (\tau_{\mu\nu}^{(2)}-2\bar G_{\mu\nu})$. Thus the coefficient $\mu_2\mu_3$ of $f^{\mu\nu}\tau^{(3)}_{\mu\nu}$ must vanish. Vanishing $\mu_2$ alone would then force $\bar \Lambda=0$, so is ruled out. Thus we conclude $\mu_3=0$.
}),
to uncover a non-trivial model we must set
$$
\Lambda=0\, ,
$$
then in turn forcing
$$
\mu_1=0\, .
$$
The only remaining algebraic $f$-terms are order three (pink): $\mu_2^2 f^3 + \mu_2\bar\Lambda f^3$. Since we must avoid setting $\mu_2=0$ (which would return us to cosmological GR),
we are forced to impose a tuning
$
\mu_2\sim \bar \Lambda
$.
From the linearized considerations of the previous Section, we can already deduce this tuning to be $$\mu_2=\frac{\bar\Lambda}6\, ,$$ in order that the FP mass obeys $m^2=\frac{2\bar\Lambda} 3$.
This value also precisely cancels the unwanted constant term in the linearized equation of motion~(\ref{leom}). To be definite, our putative PM model has equation of motion
\begin{eqnarray}\label{nlPM}
G_{\mu\nu} = \frac{\bar \Lambda}3\,  \big(f_{\mu\rho}-g_{\mu\rho} f\big) f^\rho_\nu + \frac{\bar \Lambda}6\  g_{\mu\nu}\big(f^2 - f_{\rho}^{\sigma} f^{\rho}_{\sigma} \big)\, .\\[-3.5mm] \nonumber
\end{eqnarray}

\noindent
This model strongly resembles the bimetric-motivated PM proposal of~\cite{PMmg1} (except that there $\bar \Lambda = \Lambda$) but differs sharply from the 
decoupling limit inspired PM conjecture of~\cite{PMmg}. (Possibly, heightened sensitivity of the  decoupling method  to  the contorsion difficulties we are about to  encounter 
might explain this discrepancy.)
At this juncture we can go no further with our index-free discussion and must perform an explicit computation of the fifth constraint to determine whether the model given by Eq.~(\ref{nlPM}) is~PM.

\section{Bianchi Identity?}

To  investigate explicitly the putative  PM Bianchi identity, we first gather some technical tools. 
The equation of
motion is now ${\cal G}_{\mu\nu}:=G_{\mu\nu}-\frac{\bar \Lambda}6 \tau^{(2)}_{\mu\nu}$. The vector constraint
is easy to compute, we find (denoting the inverse $f$-bein by $\ell^\mu{}_m$)
\begin{equation}\label{divdiv}
0\!=\ell _\mu^\nu {\cal C}_{\nu}\! := \ell _\mu^\nu \nabla^\rho {\cal G}_{\rho\nu}\!=-\frac{\bar \Lambda} 3\big(f^{\nu\rho} K_{\nu\rho\mu} - f K_\nu{}^\nu{}_\mu + f_\mu^\rho K_\nu{}^\nu{}_\rho\big) .
\end{equation}
Here the contorsion $K$ is defined by the difference of dynamical and background spin connections 
\begin{equation}\label{contorsion}
K_\mu{}^m{}_n:=\omega(e)_\mu{}^m{}_n-\omega(f)_\mu{}^m{}_n\, .
\end{equation}
It allows us to relate dynamical and background Riemann tensors
\begin{eqnarray*}
R_{\mu\nu}{}^{mn}&=&\bar W_{\mu\nu}{}^{mn} + \frac{2\bar\Lambda}3\,  f_{[\mu}{}^m f_{\nu]}{}^n\\[2mm]
&+&2\nabla_{[\mu} K_{\nu]}{}^{mn} - 2K_{[\mu}{}^m{}_r K_{\nu]}{}^{rn}\, .
\end{eqnarray*}
Thus, tracing the Einstein tensor with $f$  as discussed in the previous Section, we find
\begin{widetext}
\begin{eqnarray}\label{fG}
f^{\mu\nu} G_{\mu\nu}&=&
f^{\mu\sigma}\bar W_{\mu\nu}{}^\nu{}_\sigma -\frac12 f \bar W_{\mu\nu}{}^{\nu\mu}+
\frac{\bar\Lambda}6\, \big(2\, f_\mu^\nu f_\nu^\rho f_\rho^\mu - 3 f f_\mu^\nu f_\nu^\mu +  f^3\big)
\\[2mm]&+&f^{\mu\rho}\big(\nabla_\mu K_\nu{}^\nu{}_\rho-\nabla_\nu K_\mu{}^\nu{}_\rho\big)-f\nabla_\mu K_\nu{}^{\nu\mu}
-f^{\mu\sigma}K_{\mu\nu\rho}K^{\nu\rho}{}_\sigma+f^{\mu\sigma}K_{\mu\rho\sigma}K_{\nu}{}^{\nu\rho}
+\frac12\, f\big(K_{\mu\nu\rho}K^{\nu\rho\mu}+K_\mu{}^\mu{}_\rho K_\nu{}^{\nu\rho}\big)\, .\nonumber
\end{eqnarray}
\end{widetext}
Recalling that all indices are moved with the dynamical metric and vierbein, observe that the terms involving the background Weyl tensor
do not vanish (its tracelessness is with respect to $\bar g_{\mu\nu}$). As the Weyl tensor is generated nowhere else, we proceed by retreating  
from Einstein to constant curvature backgrounds by setting $\bar W_{\mu\nu}{}^{mn}=0$. This does not augur well for the putative PM model, 
since linear PM fields are known to propagate in Einstein backgrounds~\cite{DNW,DJ}.

The next task is to cancel the terms cubic in~$f$. There a temporary victory is won since
$$
 f^{\mu\nu}\tau_{\mu\nu}^{(2)}= 2\, f_\mu^\nu f_\nu^\rho f_\rho^\mu -3\, f f_\mu^\nu f_\nu^\mu +  f^3\, ,
$$
which implies (thanks to the PM tuning of $\mu_2$ to $\bar \Lambda$) that $f^{\mu\nu} {\cal G}_{\mu\nu}$ now equals the last line of~(\ref{fG}).
Those terms  involve double derivatives of the metric which can be canceled against the divergence of the vector constraint Eq.~(\ref{divdiv}) so that
\begin{equation}\label{fifth}
\begin{split}
0\ =\ {\cal C}:=&\nabla_\mu\big(\ell^{\mu\nu}\nabla^\rho{\cal G}_{\rho\nu}\big)+\frac{\bar \Lambda}3 f^{\mu\nu} {\cal G}_{\mu\nu}=\\[2mm]
& -\frac{\bar \Lambda} 3\ \Big\{  \nabla^\mu f^{\nu\rho} \big(K_{\rho\nu\mu} - g_{\nu\rho} K_\sigma{}^\sigma{}_\mu + g_{\mu\rho} K_\sigma{}^\sigma{}_\nu\big)\\[2mm]
&\ \ \  \  +\ f^{\mu\sigma}K_{\mu\nu\rho}K^{\nu\rho}{}_\sigma+f^{\mu\nu}K_{\mu\nu\rho}K_{\sigma}{}^{\sigma\rho}\\[2mm]
&\ \  \ \ -\ \frac12\, f\big(K_{\mu\nu\rho}K^{\nu\rho\mu}+K_\mu{}^\mu{}_\rho K_\nu{}^{\nu\rho}\big)\Big\}\, .
\end{split}
\end{equation}

Assuming the right hand side does NOT vanish identically, it is a constraint  (since there are no double derivatives on the metric).
Its identical vanishing is  the acid PM test!
For this test, we may employ the vector constraint~(\ref{divdiv}) 
since that would only amount to modifying the form of the putative Bianchi identity. This allows us to replace $f^{\mu\nu} K_{\mu\nu\rho}=f K_\mu - f_\mu^\nu K_\nu$
(where $K_\nu:=K_\mu{}^\mu{}_\nu$). Collecting terms and converting the $\nabla f$ term in~(\ref{fifth}) to contorsions we now face the question:
\begin{equation*}
0\ \stackrel?\equiv\ \frac12\,  f\,  K^\mu K_\mu -f_\sigma^\rho\,  K_{\mu\nu\rho} \big(K^{\nu\mu\sigma} - K^{\sigma\mu\nu}\big)-\frac12 f\,  K_{\mu\nu\rho} K^{\nu\rho\mu}\, . 
\end{equation*}
Here we may make use of any identities for the contorsion
that follow from symmetry of $f$; see Eq.~(\ref{fsymm}). A covariant derivative of that relation yields 
$$
0=K_{\rho}{}^m{}_n f_{[\mu}{}^n e_{\nu]m}-(\Gamma(g)-\Gamma(\bar g))_{\rho}{}^\sigma{}_{[\mu} f_{|\sigma|}{}^m e_{\nu]m}\, .
$$ 
Taking the  totally antisymmetric part of the above removes the difference of Christoffels term so that
$$
0=K_{[\mu\nu}{}^\sigma f_{\rho]\sigma}\, .
$$
This allows one further simplification, yielding the final query
\begin{equation}\label{zero?}
0\ \stackrel?\equiv\  \frac12 \, f\,  K^\mu K_\mu -  f^\rho_\sigma K_{\mu\nu\rho}  K^{\mu\nu\sigma}-\frac12  f K_{\mu\nu\rho} K^{\nu\rho\mu}\, . 
\end{equation}
To be absolutely certain that we are not missing some (unlikely) cancellations, we 
evaluate Eq.~(\ref{zero?}) using a solution to the vector constraint~(\ref{divdiv}) (but {\it not} of the full field equations). For that, we consider an ansatz 
$$
ds^2=-dt^2 +e^{2Mt}\, \Big(\alpha dx^2 + dy^2 + \frac{dz^2}{\alpha}\Big)
$$
for the dynamical metric in the background de Sitter coordinates
$$
d\bar s^2=-dt^2 + e^{2Mt} \Big(dx^2 + dy^2 + dz^2\Big)\, ,
$$ 
where $M^2:=\frac{\bar \Lambda}{3}$. The exact physical properties of the above ansatz are irrelevant here, we are merely verifying that no {\it identity}
vanquishes the quantity in~(\ref{zero?}). It is not difficult to verify that this ansatz obeys the vector constraint~(\ref{divdiv}) but returns 
$
{\cal C}=2M^4 \, \frac{(\alpha-1)^2}{\alpha} 
$
for the putative Bianchi identity. In other words, ${\cal C}$ is a constraint, and cannot be improved to a Bianchi identity. Despite the 
slew of algebraic cancellations achieved by the PM tuning, it did not suffice to find an identity. There is no new scalar gauge invariance removing the
zero helicity mode, hence no nonlinear PMG.

\section{Acausality}

Having dismissed the possibility of self-interacting  non-linear PM, we can apply our results to study causality of models with mass terms of type $\tau_{\mu\nu}^{(2)}$.
The results of the previous Section and~\cite{DWa} demonstrate that models 
\begin{equation}\label{12}
{\cal G}_{\mu\nu}:=G_{\mu\nu}-\Lambda g_{\mu\nu} -\mu_1\tau_{\mu\nu}^{(1)}-\mu_2 \tau_{\mu\nu}^{(2)}=0\, ,
\end{equation}
propagate five degrees of freedom for {\it all} parameter values $(\Lambda,\mu_1,\mu_2)$. Moreover the
five constraints responsible for this behavior are
\begin{widetext}
\begin{eqnarray}
{\cal C}_\nu&:=&\nabla^{\mu}{\cal G}_{\mu\nu}\ =\ -\Big[\mu_1 K^\mu 
+2\,  \mu_2  \big(f^{\rho\sigma} K_{\rho\sigma}{}^\mu - f K^\mu + f^{\mu\rho} K_\rho\big)\Big] f_{\mu\nu}\, ,\nonumber\\[3mm]
{\cal C}\ &:=&\nabla_\rho \big(\ell^{\rho\nu}\nabla^{\mu}{\cal G}_{\mu\nu})-\Big(\frac{1}2\, \mu_1 g^{\mu\nu}-2\, \mu_2 f^{\mu\nu}\Big) \, {\cal G}_{\mu\nu}\nonumber\\[2mm]
&=&2\, \mu_1\Lambda-\Big(\frac32 \, \mu_1^2 +2\, \mu_2\Lambda \Big)f
+3\, \mu_1\mu_2 \big(f^2-f_{\mu}^\nu f^{\mu}_\nu\big)
-2\, \mu_2^2\, \big(f_\mu^\nu f_\nu^\rho f_\rho^\mu -\frac32 f f_\mu^\nu f_\nu^\mu + \frac12 f^3\big)
\nonumber\\[2mm]&+&\Big[\frac12 \, \mu_1\,   e^\mu{}_n
+2\, \mu_2\big(f^\mu{}_n  -\frac12 \, f \, e^\mu{}_n \big)\Big] \, e^\nu{}_m \bar R_{\mu\nu}{}^{mn}
-\frac12 \, \mu_1 \big(K_{\mu\nu\rho}K^{\nu\rho\mu}+K_\mu K^\mu\big)\nonumber\\[2mm]
&-&2\, \mu_2 \Big[ f^\rho_\sigma K_{\mu\nu\rho}\big(K^{\nu\sigma\mu}-K^{\sigma\nu\mu}\Big)
+f^{\mu\nu}K_\mu K_\nu 
+f^{\mu\nu} K_{\mu\nu\rho} K^\rho
- \frac12\, f\big(K_{\mu\nu\rho}K^{\nu\rho\mu}+K_\rho K^{\rho}\big)\Big]\, .\nonumber \\ && \phantom{opera}  \label{cons}
\end{eqnarray}
\end{widetext}

We are now ready to study characteristics. We suppose that the dynamical metic suffers a leading discontinuity at
two derivative order across the characteristic surface~$\Sigma$
\begin{equation}\label{metS}
\big[\partial_\alpha\partial_\beta g_{\mu\nu}\big]_\Sigma=\xi_\alpha\xi_\beta\gamma_{\mu\nu}\, .
\end{equation}
Our task is to search for pathological characteristics with timelike normal 
$$
\xi^\mu g_{\mu\nu} \xi^\nu < 0\, ,
$$
with respect to the metric $g_{\mu\nu}$.
Since there is a background metric, one could also consider causal structures with respect to~$\bar g_{\mu\nu}$ and would  encounter exactly the same acausality difficulty as the one we present here. However, 
since $g_{\mu\nu}$ is the metric which couples to matter's stress tensor as well as governing the good causality properties of the leading helicity~$\pm 2$ Einstein modes,  we study it. In general acausal characteristics are ultimately associated with a breakdown of positivity of equal time commutators~\cite{JS} and thus signal inconsistency of the theory. 

We lose no generality by taking $\xi^2=-1$. Also, the metric discontinuity~(\ref{metS}) implies the leading vierbein discontinuity
$$
\big[\partial_\alpha\partial_\beta e_{\mu}{}^m\big]_\Sigma=\xi_\alpha\xi_\beta {\cal E}_\mu{}^m\, ,
$$
where the leading discontinuity in the relation $e_{\mu}{}^m e_{\nu m}=g_{\mu\nu}$ implies
$$
2\, {\cal E}_{\mu\nu}=\gamma_{\mu\nu} + a_{\mu\nu}\, ,$$
with $a_{\mu\nu}=-a_{\nu\mu}$. 

Absence of acausal characteristics would hold if the algebraic set of conditions following from the leading discontinuity in: (i) the equation of motion~(\ref{12}), (ii) the constraints~(\ref{cons})
and (iii) the symmetry condition~(\ref{fsymm}), forces $\gamma_{\mu\nu}=0=a_{\mu\nu}$ when $\xi^2=-1$. Any causality violations of course appear in lower helicity sectors
because the leading discontinuity of the equation of motion is that of Einstein's theory:
$$
\xi^2 \gamma_{\mu\nu} - \xi_\mu\,  \xi.\gamma_\nu - \xi_\nu\,  \xi.\gamma_\mu + \xi_\mu\xi_\nu\, \gamma=0\, .
$$
\vspace{0mm}

\noindent
This implies that the transverse part $\gamma_{\mu\nu}^\perp=0$. In what follows we will decompose tensors with respect to the (unit) timelike vector~$\xi_\mu$ according to 
$$
\begin{array}{cclr}
V_\mu&\!\!\!:=\!\!&V^\perp_\mu -\xi_\mu \xi.V\, ,\\[2mm]
\!\!S_{\mu\nu}&\!\!\!:=\!\!&
S^\perp_{\mu\nu}\!-\xi^\noperp_\mu S_\nu^\perp -\xi^\noperp_\nu S_\mu^\perp \!+ \xi_\mu\xi_\nu\,  \xi.\xi.S\, ,&\hspace{-1mm}\big(S_\mu:=\xi.S_\mu\big) ,\\[2mm]
A_{\mu\nu}&\!\!\!:=\!\!&A^\perp_{\mu\nu}+\xi^\noperp_\mu A_\nu^\perp -\xi^\noperp_\nu A_\mu^\perp\, ,&\hspace{-13mm}\big(A^\perp_\mu:=A_{\mu\nu} \xi^\nu\big) ,\\
\end{array}
$$
where $V$, $S$ and $A$ denote a vector, and  symmetric and antisymmetric tensors, respectively.

At this juncture, of the sixteen components of $\gamma_{\mu\nu}$, and $a_{\mu\nu}$, the ten encoded by
$\gamma_\mu^\perp$ (three), $\xi.\xi.\gamma$ (one) $a_{\mu\nu}^\perp$ (three) and $a_\mu^\perp$ (three), 
remain. The discontinuity in the symmetry relation~(\ref{fsymm}) gives six homogeneous conditions on these:
\begin{widetext}
\begin{equation*}
f^\perp_{[\mu}{}^\rho a_{\nu]\rho}^\perp-f_{[\mu}^\perp \Big\{a_{\nu]}^\perp+ \gamma_{\nu]}^\perp\Big\} \ = \ 0\ = \  
 f^\perp_\rho a^\perp_\nu{}^\rho  +f^\perp_\mu{}^\rho \big(a^\perp_\rho - \gamma^\perp_\rho\big) 
+f^\perp_\mu \xi.\xi.\gamma
-\xi.\xi.f\, \big(a^\perp_\mu +f^\perp_\mu \big)\, .
\end{equation*}
\end{widetext}
At very best, at this point only four combinations of the ten variables $(\gamma_\mu^\perp, \xi.\xi.\gamma, a_{\mu\nu}^\perp, a_\mu^\perp)$
are left. Thus we need four more conditions to establish the absence of acausal characteristics. These can only come from the four constraints~(\ref{cons})
(further constraints would anyway destroy the DoF count). The leading discontinuity in the constraints is given by the metric derivatives in the contortions and thus 
proportional to $[\partial_\alpha K_{\mu\nu\rho}]_\Sigma$. This quantity is easily computed
\begin{equation*}
\begin{split}
2\,\xi^\alpha  [ &\partial_\alpha   K_{\mu\nu\rho}]_\Sigma\\[1mm]&
=\xi_\nu {\cal E}_{\rho\mu}-\xi_\rho {\cal E}_{\nu\mu}
-\xi_\mu {\cal E}_{\nu\rho}+\xi_\nu {\cal E}_{\mu\rho}
+\xi_\mu {\cal E}_{\rho\nu}-\xi_\rho {\cal E}_{\mu\nu}\\&
= - \xi_\mu a^\perp_{\nu\rho}-2\, \xi_\mu \xi_{[\nu} \Big\{a^\perp_{\rho]}+\gamma^\perp_{\rho]}\Big\}\, .
\end{split}
\end{equation*}
Thus the discontinuity in the constraints gives four homogeneous linear conditions on the six quantities $( a_{\mu\nu}^\perp, a_\mu^\perp+\gamma_\mu^\perp)$.
To summarize we have the following linear system of ten equations in ten unknowns:

\vspace{2mm}
\begin{tabular}{|c|c|}\hline
\multirow{2}{*}{variables} &  \ \ homogeneous\ \  \\ & \ \ conditions\ \ \\ \hline &\\[-3mm]
$a^\perp_{\mu\nu}$, $a^\perp_\mu+\gamma^\perp_\mu$ & 7\\[2mm]
\ \ $a^\perp_{\mu\nu}$, $a^\perp_\mu+\gamma^\perp_\mu$, $a^\perp_\mu-\gamma^\perp_\mu$,  $\xi.\xi.\gamma$ \ \ & 3\\[2mm]
 \hline
\end{tabular} 
\vspace{2mm}

\noindent
Evidently, from the first line of the table, seven homogeneous conditions on the six variables $(a^\perp_{\mu\nu}, a^\perp_\mu+\gamma^\perp_\mu)$ will generically force these to vanish
which, by itself, bodes well for causality (non-generic conditions that do not  kill $(a^\perp_{\mu\nu}, a^\perp_\mu+\gamma^\perp_\mu)$ already correspond to acausal characteristics). But there are only three conditions on the four remaining variables ($a^\perp_\mu-\gamma^\perp_\mu,  
\xi.\xi.\gamma$), which means that some combination thereof does not vanish, so there are acausal characteristics.

\section{Conclusions}

We have demonstrated that none of the ghost-free, $f$-$g$  massive gravity models of~\cite{deRham, HR}
 exhibits partial masslessness~\footnote{For one model (see Eq.~(\ref{nlPM}) this failure involves only
 terms in the fifth constraint mades from squares of contorsions in constant curvature backgrounds (but also Weyl terms in Einstein ones).}.
The same terms
are responsible for acausality of  ghost-free, $f$-$g$  massive gravity models~\footnote{Strictly speaking, absent an explicit computation of the covariant constraint for the $\tau^{(3)}_{\mu\nu}$ mass term, these
models could conceivably survive, but it seems far more likely that the  form of the fifth constraint and the generality of our characteristics computation will rule that model out too.}. These results are consistent with earlier order by order analyses of PM self-interactions~\cite{Zinoviev} that claimed no consistent self-couplings existed beyond (as usual, safe) cubic order~\footnote{In fact, since $s=2$, $d=4$ PM is gauge invariant, propagates on light cone~\cite{PMlight}, is conformally~\cite{PMconf} and duality invariant~\cite{PMdual}, and couples consistently to charged matter, it might make more sense to search for non-abelian Yang--Mills-like interactions.}. A conformal gravity inspired PM study reached the same conclusion~\cite{DJ}.  The old lesson (first learnt in a  charged massive $s=3/2$ context~\cite{JS}) is again at play here,
healthy DoF counts alone need not imply physical consistency.

\begin{acknowledgments}
We thank Nemanja Kaloper for discussions.
 S.D. was supported in part by NSF PHY- 1064302 and DOE DE-FG02-164 92ER40701 grants. 
\end{acknowledgments}


\end{document}